\documentstyle[aps,prl,epsfig]{revtex}

\begin{document}

\twocolumn[\hsize\textwidth\columnwidth
\hsize\csname@twocolumnfalse\endcsname
\draft

\title{Nuclear spin driven quantum relaxation in
LiY$_{0.998}$Ho$_{0.002}$F$_{4}$}

\author{R. Giraud$^1$, W. Wernsdorfer$^1$, A.M. Tkachuk$^2$, D. Mailly$^3$,
and B. Barbara$^1$}

\address{$^1$ Laboratoire de Magn\'etisme Louis N\'eel, CNRS, BP166, 38042
Grenoble Cedex-09, France\\
$^2$ All-Russia Scientific Center ``S.I. Vavilov State Optical Institute'',
199034 St. Petersburg, Russia\\
$^3$ Laboratoire de Photonique et de Nanostructures, CNRS, 196 Av. H.
Ravera, 92220 Bagneux, France\\}

\date{\today}

\maketitle

\begin{abstract}

Staircaselike hysteresis loops of the magnetization of a
LiY$_{0.998}$Ho$_{0.002}$F$_{4}$ single crystal are observed at subkelvin
temperatures and low field sweep rates. This behavior results from quantum
dynamics at avoided level crossings of the energy spectrum of single
Ho$^{3+}$ ions in the presence of hyperfine interactions. Enhanced quantum
relaxation in constant transverse fields allows the study of the relative
magnitude of tunnel splittings. At faster sweep rates,
nonequilibrated spin--phonon and spin--spin transitions, mediated by weak
dipolar interactions, lead to magnetization oscillations and additional steps.

\end{abstract}

]

\narrowtext

The problem of quantum dynamics of a two-level system coupled to an
environment (boson or fermion bath) is at the core of mesoscopic physics
\cite{Leggett87}. We show that the new field of ``mesoscopic magnetism'', which
studies the tunneling of large magnetic moments in the presence of phonons and
spins, is not limited to molecular complexes and nanoparticles, but it can be
extended to other systems such as rare-earth ions. After the first studies on
large spin molecules Mn$_{12}$-ac \cite{Novak95BBjmmm95,LucFriedman96} and
Fe$_{8}$ \cite{Fe8}, the role of the spin bath on the tunnel mechanism was
shown \cite{ProkStampGarg,ProkStamp,Werns00,jmmm200,Igor}. In particular,
quasistatic fields due to dipolar interactions between molecules lead to a
distribution of internal fields, and field fluctuations, essentially of
nuclear spins, give homogeneous level broadening allowing the restoration of
tunneling in a finite energy window, at low temperature; this broadening being
much larger than the phonon one, it is more relevant to induce tunneling. This
mechanism is efficient unless all nuclear spins of the molecule are frozen,
which occurs only below the mK scale. In low spin molecules, large tunneling
gaps favor spin--phonon transitions. Although the hyperfine induced level
broadening is the same as in large spin molecules, the phonon bath becomes as
important as the spin bath \cite{V15}. In all these cases, the role of field
fluctuations was clearly evidenced. 

This description is for the relatively weak hyperfine interactions of
Mn$_{12}$ or Fe$_{8}$ molecules, and therefore for incoherent nuclear spin
fluctuations. The question as to what really happens when an electronic moment
tunnels, while it is strongly coupled to its nuclear spin, has not yet a
clear answer. Contrary to the 3$\it d$ group, hyperfine interactions are very 
large in 4$\it f$ elements. Diluted rare-earth ions in a nonmagnetic 
insulating single
crystal are therefore very suitable to study the possible entanglement of
nuclear and electronic moments, when tunneling occurs. Our choice was the
weakly doped rare-earth fluoride series LiY$_{1-x}$R$_{x}$F$_{4}$, in which
high quality single crystals are mainly investigated for applications in
high-power laser diodes \cite{Tkachuk78}. Note that EPR spin-echo of magnetic
tunneling states have already been observed in a $1\%$ Dy-doped
crystal\cite{Bellessa98}. At higher concentrations, these crystals were used
for phase transition studies of dipolar ordered magnets
\cite{Hansen75Beauvillain}. Among them, the holmium doped fluoride is a
random, dipolar coupled system with an Ising ground state doublet
($g_{\rm eff}\approx13.3$ \cite{Magarino76}, see also \cite{Mennenga84} and
references therein) and a pure isotope $I=7/2$ nuclear spin. The magnetic
properties of the Ising ferromagnet LiHoF$_{4}$ and spin glass
LiY$_{0.833}$Ho$_{0.167}$F$_{4}$ have been studied by ac susceptibility
\cite{Reich}. In particular, enhanced quantum fluctuations, leading to a
cross-over to a quantum spin glass, were evidenced in an applied transverse
field \cite{Aeppli91}. 

\begin{figure}
\centerline{\epsfxsize= 8.1 cm \epsfbox{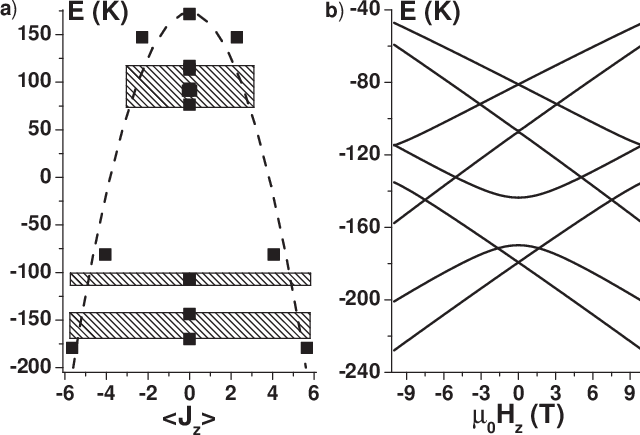}}
\caption{(a): Energy levels vs average value of $J_z$, in zero applied
field, showing an Ising ground state doublet and a first excited singlet at
$\approx9.5$~K above. (b): Low-energy part of the Zeeman diagram. The
first excited state ($\approx25$~K below the next excited $\Gamma_{2}$
singlet) defines an energy barrier of $\approx9.5$~K hindering the magnetic
moment reversal.}    
\label{fig1}   
\end{figure}

\begin{figure}     
\centerline{\epsfxsize= 8.1 cm \epsfbox{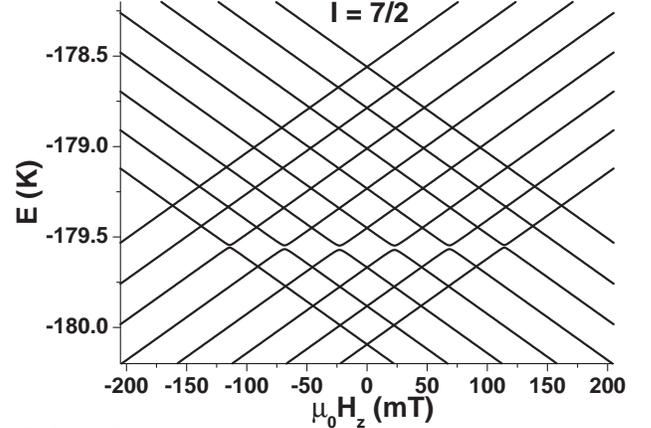}}  
\caption{Splitting of the electronic ground state doublet by the hyperfine
interaction ($A_J/k_{\rm B}\approx38.6$~mK; see below). The level crossings
occur for resonant values of the longitudinal field $H_n$ ($-7\le n\le 7$).
Some are avoided level crossings and hyperfine levels repulsion is then
induced by the electronic level repulsion in the excited states shown in
Fig.~\ref{fig1}(b).}         
\label{fig2}  
\end{figure}

In this letter, we investigate a single crystalline $0.2\%$ holmium doped
LiYF$_{4}$ at subkelvin temperatures. The isolated magnetic moments are
weakly coupled by dipolar interactions ($\mu_0 H_{\rm dip} \sim$~ few mT) so
that this very diluted insulator should exhibit a nearly single ion quantum
behavior, in which we are interested in continuity with our studies on
molecular magnets. The crystal has a tetragonal scheelite structure with a
C$_{4h}$ space symmetry group ($I4_1/a$), and the point symmetry group at
Ho$^{3+}$ sites is S$_{4}$, almost equivalent to D$_{2d}$ (for LiHoF$_{4}$,
unit cell parameters are $a=b=5.175$~$\AA$ and $c=10.74$~$\AA$
\cite{Thoma61}). Because of a very strong spin--orbit coupling, each magnetic
ion of $^{165}$Ho is characterized by its $J=8$ ground state manifold
(g$_J=5/4$), split by crystal field effects. These last give rise to a large
uniaxial magnetic anisotropy, {\it i.e.}, a high energy barrier hindering the
magnetic moment reversal. However, we will see that quantum fluctuations due to
significant transverse anisotropy terms drastically reduce this barrier. This
effect was very much weaker in Mn$_{12}$-ac ($\approx10\%$ barrier reduction
\cite{jmmm200}). Using the $\vert J,M>$ basis and D$_{2d}$ symmetry, the
approximate Hamiltonian including hyperfine interaction writes 
\begin{eqnarray}     
H = H_{\rm crystal~field} + H_{\rm Zeeman} + H_{\rm hyperfine}   
\end{eqnarray}  
with 
\begin{eqnarray}   
H_{\rm crystal~field} = \alpha_J B{^0_2}O{^0_2} \nonumber
\end{eqnarray}

\begin{eqnarray}
 + \beta_J (B{^0_4} O{^0_4}+B{^4_4}O{^4_4}) + \gamma_J
(B{^0_6} O{^0_6}+B{^4_6} O{^4_6}),   \nonumber
\end{eqnarray}

\begin{eqnarray}
H_{\rm Zeeman} = -g_J\mu_{\rm B}\vec{J}\cdot\vec{H}, ~ 
H_{\rm hyperfine} = A_J \vec{J}\cdot\vec{I}. \nonumber 
\end{eqnarray}
The $\alpha_J$, $\beta_J$, $\gamma_J$, and $O{^m_l}$ are Stevens' coefficients
and equivalent operators \cite{Stevens}. Exact diagonalization of the
$136$-dimensional Hamiltonian (1) was performed, using a set of crystal
field parameters obtained by high-resolution optical spectroscopy:
$B{^0_2}=273.9$~K, $B{^0_4}=-97.7$~K, $B{^0_6}=-6.5$~K, $B{^4_4}=-1289.1$~K,
$B{^4_6}=-631.6$~K \cite{Tkachuk78}. $A_J$ was taken as a fitting parameter
of the measured resonances (see below). $\it J$ mixing, Jahn-Teller effect and
nuclear quadrupole interaction are assumed to be negligible. We first show the
results with $A_J=0$ in Fig.~\ref{fig1}.
The eigenstates transform as one of the four irreducible
representations $\Gamma_{1,2,3,4}$ of the S$_4$ point group. Significant
transverse crystal field terms $B^4_4 O{^4_4}$ and $B^4_6 O{^4_6}$ mix free
ion states, with $\Delta M=\pm4$, so that eigenvectors are linear combinations
either of $\vert\pm7>,\vert\pm3>,\vert\mp1>$ and $\vert\mp5>$ for 
$\Gamma_{3,4}$; $\vert\pm6>,\vert\pm2>,\vert\mp2>$ and $\vert\mp6>$ for
$\Gamma_{2}$; or $\vert\pm8>,\vert\pm4>,\vert0>,\vert\mp4>$ and $\vert\mp8>$
for $\Gamma_{1}$. In Fig.~\ref{fig1}a), the calculated low-lying states
within the $^5I_8$ multiplet show a $\Gamma_{34}$ Ising ground state doublet
noted as $\vert \psi_1^{\pm}>$, while the first excited state, a $\Gamma_2$
singlet, stands at $\approx 9.5$~K above (direct measurements
give $\approx 10\pm1$~K \cite{Mennenga84}). Fig.~\ref{fig1}(a) also shows how
the expected large barrier $\sim10^2$~K is shortcut by large tunneling gaps
due to transverse crystal field terms (emphasized by shaded areas between
singlets belonging to the same representation). A strong electronic level
repulsion in the low-lying excited $\Gamma_{2}$ states is clearly shown in
Fig.~\ref{fig1}(b). This defines the energy barrier the magnetic moment has to
overcome in order to reverse its polarization. At very low temperatures, the
system should be equivalent to a two-level system with an energy barrier of
$\approx10$~K. Actually, this picture is strongly modified when intraionic
dipolar interactions are taken into account ($A_J\ne0$). They lead
to a more complex diagram in the electronic ground state, showing several level
crossings for resonant values $H_n$ ($-7\le n\le 7$) in Fig.~\ref{fig2}. The
transverse hyperfine contribution $\frac{1}{2} A_J (J_+I_- + J_-I_+)$ induces
some avoided level crossings between $\vert \psi_1^-,I_{z1}>$ and $\vert
\psi_1^+,I_{z2}>$, with $\Delta I=\vert  I_{z2}-I_{z1}\vert$, only when
$\Delta I/2$ is an odd integer so that the two electronic low-lying states
$\vert \psi_1^{\pm}>$ are coupled through nondegenerated excited electronic
levels.          

\begin{figure}
\centerline{\epsfxsize= 8.1 cm \epsfbox{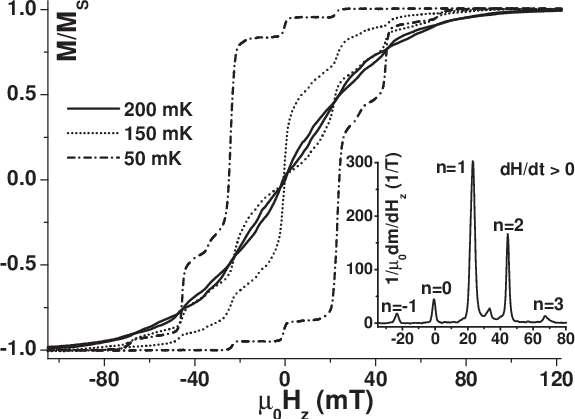}} 
\caption{Hysteresis loops for $v=0.55$~mT/s and at different temperatures.
Inset: derivative of the magnetization normalized to $M_{\rm S}$, d$dm/dH$, at
$T=50$~mK and for $v=0.55$~mT/s.}    
\label{fig3}
\end{figure}

Magnetic measurements were made at $0.04<T<1$~K and for $\mu_0H<2$~T,
with a micro-SQUID magnetometer \cite{Werns97} allowing field sweep rates up to
1~T/s. The crystal is first saturated in a large positive field applied along
the $c-$axis $\mu_0H_{\rm sat}\approx0.3$~T, and then the field $H_z$ is swept
between $\pm H_{\rm sat}$.  At slow field sweep rates, an $\it isothermal$
process occurs, leading to staircaselike hysteresis loops at $T\le200$~mK, as
shown in Fig.~\ref{fig3}. These well defined steps come from quantum
relaxation at avoided level crossings. At $T=50$~mK, the lowest
energy level is mainly populated and magnetization steps are observed for
$-1\le n\le 3$. The ground and excited tunnel splittings at $H_z=0$, being due
to very small perturbations such as internal transverse fields, should be very
similar. Therefore, quantum tunneling in zero field is mainly associated with
the dynamics of the lowest avoided level crossing (the first excited crossing
is at $\Delta E/k_{\rm B} = g_{\rm eff}\mu_{\rm B}H_1/k_{\rm B} \approx205$~mK,
assuming $g_{\rm eff}\approx13.3$ \cite{Magarino76}). The amplitude of the
next step,  the resonance $n=1$ at $\mu_0 H_1=23$~mT, is about ten times
larger, suggesting a larger tunnel splitting $\Delta$. Indeed, Fig.~\ref{fig2}
shows that the hyperfine induced tunnel splitting of the third excited avoided
level crossing is large enough to render the barrier transparent ($\Delta
\approx25$~mK). The relaxation time is thus simply given by thermal activation
$\tau…= \tau_0 \exp(2\Delta E/k_{\rm B}T)$, with a long $\tau_0$ because 
spin--lattice
relaxation time $T_1$ can be hours at very low temperatures and/or as a
result of internal fields fluctuations.   
In the same trend, the measured magnetization step ratio ${\Delta
M(n=1)}/{\Delta M(n=-1)} \approx25$ at $T=50$~mK is approximately equal to the
Boltzmann ratio, which confirms that thermally activated quantum relaxation
occurs at $H_z\ne0$. The barrier, essentially transparent due to this large
splitting, becomes again finite out of resonance. 
\begin{figure}  
\centerline{\epsfxsize= 8.1 cm \epsfbox{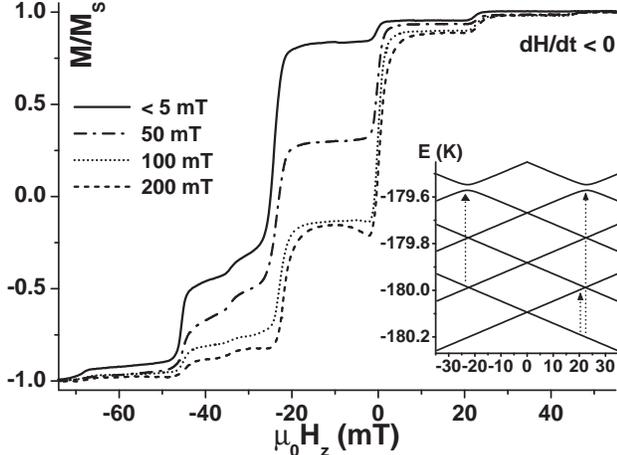}}
\caption{Hysteresis loops in a constant transverse field at $T=50$~mK and for
$v=0.55$~mT/s. A transverse field enhances the quantum fluctuations in zero
longitudinal applied field leading to a larger magnetization step. Inset:
details of the Zeeman diagram around zero field. Thermally activated tunneling
shows two possible channels over the first and the third, more efficient,
excited avoided level crossings.}     
\label{fig4} 
\end{figure} 
The quantum relaxation is strongly enhanced by a constant transverse field,
as a result of an increase of the tunnel splittings (see Fig.~\ref{fig4}). In
zero longitudinal field, the small tunnel splittings rapidly increase and
hysteresis vanishes. A saturation of the magnetization at $M\approx0$ is
observed in transverse fields larger than $100$~mT, when the barrier is nearly
transparent, and the small ``overshot'' with an oscillation in $M$ may be due
to spin--phonon transitions. As expected for a large tunnel splitting,
sensitivity to a small transverse field is very weak for the resonance $n=-1$.
The inset in Fig.~\ref{fig3} shows $dm/dH$ at $T=50$~mK. The width of the
resonant transitions is about $\mu_0 \Delta  H=2-3$~mT which is expected from
dipolar broadening. Similarly to molecular magnets, quasi-static fields due to
dipolar interactions lead to a distribution of internal fields whereas field
fluctuations, essentially of F$^-$ nuclear spins, give homogeneous level
broadening. 

\begin{figure}
\centerline{\epsfxsize= 8.1 cm \epsfbox{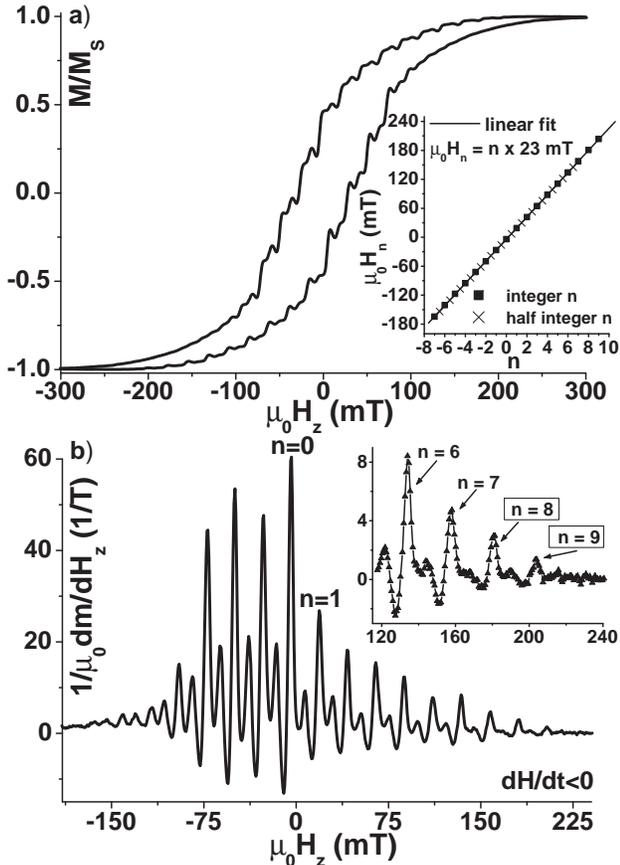}}
\caption{(a): Hysteresis loops at $T=50$~mK and for $v=0.3$~T/s. Several
magnetization steps are observed for resonant values of the applied field
$\mu_0 H_n\approx n\times 23$~mT [see inset; $H_n$ values are deduced from
Fig.~\ref{fig5}(b)]. (b): Derivative of the loop shown in (a) for a
decreasing field. The two additional measured steps shown in the inset, for
$n=8$ and $n=9$, are associated with cross-spin relaxation only.}   
\label{fig5}     
\end{figure}

A hysteresis loop measured at $T=50$~mK for a much faster field sweep rate
($v=0.3$~T/s) is shown in Fig.~\ref{fig5}(a). A succession of equally spaced
large and weak magnetization steps occur at fields $H_n$, with $-14\le2n\le14$.
The larger ones, with integer $n$, are associated with several equally spaced
level crossings and the smaller steps, with half integer $n$, fall just in
between when the levels are equally spaced (see Fig.~\ref{fig2}). $dm/dH$ is
used to determine the $H_n$ values plotted in Fig.~\ref{fig5}(a) inset. From 
the
slope, we accurately obtain $\mu_0 H_n= n\times23$~mT. The electronic ground
doublet is thus split by hyperfine interaction in eight doublets over an
energy range of about $1.44$~K. We deduce $A_J/k_{\rm B}\approx38.62$~mK, to be
compared to $A_J/k_{\rm B}\approx40.95$~mK \cite{Magarino76}. The observed
hysteresis loops depend sensitively on sample thermalization, showing that the
spin--phonon system is not at equilibrium with the cryostat, leading to a
phonon bottleneck \cite{V15,ABScott62Shapira91,Curtis}. At a fast field sweep
rate $v=0.3$~T/s, the system enters such a regime at $T\approx 1$~K (moderate
sample thermalization) showing hysteresis without any magnetization steps down
to $T\approx600$~mK. When the field is swept back and forth, a stationary
regime occurs and hysteresis loops become nearly temperature-independent below
a temperature $T_c(v)$ depending on sample thermalization ($T_c\approx200$~mK
for $v=0.3$~T/s). Below $T\approx600$~mK, a nearly $\it adiabatic$ process
occurs, due to a much longer spin--lattice relaxation time $T_1$. The spin 
system becomes more and more isolated from the phonon bath, and energy exchange
between electronic and nuclear spins is only possible at fields $H_n$.
Equilibrium within the spin system is due to either quantum fluctuations at 
avoided level crossings (integer $n$) or to spin--phonon transitions and/or 
cross-spin relaxation, allowed by weak dipolar interactions, when energy 
levels are almost equally spaced (integer and half integer $n$)
\cite{Curtis,Bloem59Hellwege68}. Spin--spin interactions allow two
additional steps for $n=8$ and $n=9$, at fields with equally spaced levels but
no level crossing [Fig.~\ref{fig5}(b) inset]. A small transverse
applied field only increases the zero-field magnetization step, showing the 
weak effect of enhanced quantum fluctuations on hysteresis loops in this
regime. Other resonances and small magnetization steps, dominated by
cross-spin relaxation, are not affected by a small transverse field, if small
enough ($\mu_0 H_T \lesssim20$~mT).  If the field sweep is suddenly
stopped, the spin--phonon system exchanges energy with the cryostat and the
magnetization relaxes toward the equilibrium curve.   
The asymmetry of the envelope (not drawn) of the peaks in Fig.~\ref{fig5}(b),
showing that spins reverse mostly after field inversion, and the absence of
constriction in the hysteresis loop near $H_z=0$, confirm the existence of
small barriers, mainly in zero field (tunneling).

In conclusion, we have shown that the quantum rotation of weakly coupled
magnetic moments in LiY$_{0.998}$Ho$_{0.002}$F$_{4}$ can be driven and
monitored by hyperfine couplings. At very low temperatures, when the field is
slowly swept from negative to positive values, the coupled electronic and
nuclear moments tunnels from $\vert \psi_1^-,I_{z1}>$ to $\vert
\psi_1^+,I_{z2}>$. In a constant transverse field, the magnetization step,
associated with incoherent tunneling at the avoided level crossing
in zero field, increases very rapidly. It saturates when the barrier is
completely transparent. Details of hysteresis loops are in excellent agreement
with the level structure of the electronic ground state doublet split by
hyperfine interaction in sixteen states. At faster field sweep rates,
additional magnetization steps are observed and attributed to cross-spin
relaxation and/or spin--phonon transitions in a phonon bottleneck regime.
Very  diluted Holmium doped LiYF$_{4}$ is thus a model system to study
tunneling of an electronic moment strongly coupled to its nuclear spin.

We are very grateful to I. Chiorescu, J.C. Vial, and A.K. Zvezdin for
discussions and to M.F. Joubert and P. Lejay for on-going collaborations.
This work has been supported by DRET, Rh\^one-Alpes, MASSDOTS ESPRIT,
MolNanoMag TMR and AFIRST.

\end{document}